\documentstyle[psfig,amssymb,amsfonts,12pt,preprint,tighten,aps]{revtex}

\begin{document}
\input psfig.sty
\title{Addendum to: Update on neutrino mixing in the early Universe}
\author{P. Di Bari}
\maketitle
\begin{center}
{\em IFAE, Universitat Aut{\`o}noma de Barcelona,  08193 Bellaterra (Barcelona), Spain} \\
{\it\small e-mail: dibari@ifae.es}
\end{center}
\date{}
\maketitle
\begin{abstract}
{In the light of the recent WMAP results we update the constraints on a class
of non standard BBN models with a simultaneous combination of 
non standard neutrino distributions and extra effective number of neutrinos
in the expansion rate.  These models can be described in terms
of the two parameters $\Delta N_{\nu}^{\rm tot}$, constrained 
by the primordial Helium abundance $Y_p$ measurement,  and $\Delta N_{\nu}^{\rho}$, 
constrained by a combination of CMB and primordial Deuterium data.  
Small deviations from standard Big Bang Nucleosynthesis are suggested. 
Different non standard scenarios can be 
distinguished by a measurement of the difference 
$\Delta N_{\nu}^{f_{\nu}}=\Delta N_{\nu}^{\rm tot}-\Delta N_{\nu}^{\rho}$. 
From the current data we estimate $\Delta N_{\nu}^{f_{\nu}}\simeq -1.4^{+0.9}_{-1.4}$,
mildly disfavouring solutions with a low expansion rate, characterized by $\Delta N_{\nu}^{f_{\nu}}=0$ and negative $\Delta N_{\nu}^{\rho}$.  
Active-sterile neutrino mixing could be 
a viable explanation only for high values of $Y_p\gtrsim 0.24$. 
The existence 
of large positive neutrino chemical potentials $\xi_i\sim 0.05$, 
implying $\Delta N_{\nu}^{\rho}\simeq 0$, would be a possible explanation of the data
within the analyzed class of non standard BBN models. Interestingly 
it would also provide a way to evade the cosmological bounds for 
`class A 3+1' four neutrino mixing models. A scenario with a decaying sterile neutrino
is also considered.}
\end{abstract}

\newpage
\section{Introduction}

In a previous paper \cite{main} (see also \cite{buda}) we showed how the new CMB measurements 
of the baryon to photon ratio, $\eta$, are able to put stringent constraints
on a large class of non standard BBN models where, together with a usual
variation of the expansion rate due to the presence of extra degrees of freedom, 
distortions of the electron neutrino distribution are also present. 
This class of models can be described in terms of two parameters \cite{ropa}. 
A first one is the usual extra effective number of neutrinos, modifying the standard 
expansion rate, $\Delta N_{\nu}^{\rho}=[\sum_{X}\,(\rho_X/\rho_0)-3]$,
where $\rho_X$ is the energy density of the $X$-particle species, including
the three ordinary neutrinos plus possible new ones, and
$\rho_0=(7\pi^2/120)\,T^4_{\nu}$ is the energy density of one standard
neutrino species.  The second one is the total extra effective number of neutrinos 
$\Delta N_{\nu}^{\rm tot}$ defined, in terms of the primordial $^4$He abundance $Y_p$, as:
\begin{equation}
\Delta N_{\nu}^{\rm tot}=
\left[Y_p^{BBN}(\eta,\Delta N_{\nu}^{\rho},\delta f_{\nu_e})-Y_p^{SBBN}\right]/0.0137 \,\, .
\end{equation} 
The difference $\Delta N_{\nu}^{\rm tot}-\Delta N_{\nu}^{\rho}$ is a quantity
that, in the class of models that we are considering, has to be entirely ascribed to the effect
of deviations of the electron electron neutrino distribution from the standard Fermi-Dirac
with zero chemical potential, $\delta f_{\nu_e}=f_{\nu_e}-f^0_{\nu_e}$.
If $\delta f_{\nu_e}=0$ then $\Delta N_{\nu}^{\rm tot}=\Delta N_{\nu}^{\rho}$
and simply \cite{lopez}:
\begin{equation}
Y_p^{BBN}(\eta,\Delta N_{\nu}^{\rho},\delta f_{\nu_e}=0)\simeq
Y_p^{SBBN}(\eta)+\gamma(\eta)\,\Delta N_{\nu}^{\rho}
\end{equation}
with $\eta$ the baryon to photon ratio in units of $10^{-10}$. 
Using the expansion given in \cite{lopez}, we calculated that 
$\gamma(\eta)\simeq 0.0137$ over the pertinent range $\eta=3.5-10$. 
The standard BBN prediction for $Y_p$ is described by the following
expansion around $\eta=5$ \cite{lopez}:
\begin{equation}
Y_p^{SBBN}\simeq 0.2466+0.01\,\ln\left(\eta/ 5\right)
\end{equation}
The presence of non zero $\delta f_{\nu_e}$ affects mainly $Y_p$, 
while its effect can be safely neglected in the Deuterium
abundance, $(D/H)$, also considering that we will be interested in small
deviations. With this approximation the $D/H$ abundance is 
described by the expression \cite{main}:
\begin{equation}\label{DH}
(D/H)^{BBN}(\eta,\Delta N_{\nu}^{\rho})\simeq
\left[3.6\cdot 10^{-5}\,\left(\eta/ 5\right)^{-\beta}\right]
\left(1+\alpha\,\Delta N_{\nu}^{\rho}\right)^{\beta\over 2} 
\end{equation}
with $\beta\simeq 1.6$ and $\alpha=(7\,r^4_{{\nu}0}/4g_{\rho 0}^{SBBN})\simeq 0.135$, 
where $r_{\nu 0}$ and $g_{\rho 0}^{SM}$ 
are, respectively, the standard 
neutrino to photon temperature ratio and the 
number of degrees of freedom at present.
With these expressions a simultaneous measurement of  
$(D/H)$, $Y_p$ and $\eta$ can be easily translated into a `measurement'
of $\Delta N_{\nu}^{\rm tot}$ and $\Delta N_{\nu}^{\rho}$.
We used in \cite{main} both high \cite{Izotov}
\footnote{We indicate $68\%$ c.l. errors for all quantities unless differently indicated.} 
\begin{equation}\label{high}
Y_p^{\rm exp}=0.244\pm 0.002
\end{equation}
and low values $Y_p^{\rm exp}=0.234\pm 0.003$,
while we used $(D/H)^{\rm exp}=(3.0\pm 0.4)\times 10^{-5}$ \cite{O'Meara}.
For $\eta$ we used the DASI and BOOMerANG result \cite{CMB} $\eta^{CMB}=6.0^{+1.1}_{-0.8}$.
From low values of Helium and assuming gaussian
errors, we obtained  $\Delta N_{\nu}^{\rm tot}=-1.05 \pm 0.25$,  
while from high values of Helium we obtained $\Delta N_{\nu}^{\rm tot}=-0.3\pm 0.2$.
Using the primordial Deuterium abundance measurement,  
from the expression (\ref{DH}), we could
estimate $\Delta N_{\nu}^{\rho}$ obtaining $\Delta N_{\nu}^{\rho}=1\pm 4 $.
These results were implying, at $3\,\sigma$ the bounds \cite{main}
$\Delta N_{\nu}^{\rm tot}< 0.3$ and $\Delta N_{\nu}^{\rho}\lesssim 13$.
In particular the bound on $\Delta N_{\nu}^{\rm tot}$ 
was used to conclude that, for negligible neutrino asymmetries, 
all four neutrino mixing models
are in disagreement with cosmology and thus ruled out. This has been then also 
confirmed by the improved solar and atmospheric neutrino data from the SNO \cite{SNO}
and SuperK \cite{SK} experiments \cite{Maltoni}. 
In the next section we will update these results in light, mainly, of the 
recent results from the WMAP experiment \cite{WMAP} and we will see how 
the data suggest possible deviations from a standard picture.

\section{Updated reference values and results}

The WMAP collaboration finds $\Omega_b\,h^2=0.0224 \pm 0.0009$ \cite{WMAP}
corresponding to:
\begin{equation}\label{eta}
\eta^{CMB}=6.15\pm 0.25
\end{equation}
This measurement is so precise that now, when estimating $\Delta N_{\nu}^{\rm tot}$,
the experimental error on $Y_p$ is dominant compared to that one on $\eta$. 
Using high values of 
$Y_p^{\rm exp }$  we find at 1 $\sigma$: $\Delta N_{\nu}^{\rm tot}=-0.35\pm 0.15$.
This means that now a 3\,$\sigma$ range is given by
\begin{equation}
-0.8< \Delta N_{\nu}^{\rm tot}< 0.1\, ,
\end{equation} 
implying a quite more stringent upper bound compared to the pre-WMAP value. 
Even using the range of values
\begin{equation}\label{Ypm}
Y_p^{\rm exp}= 0.238 \pm 0.002 \pm 0.005\,\, ,
\end{equation}
that is a compromise between low and high values and takes into account the
discrepancy as a sistematic uncertainty \cite{Fields}, we find
\begin{equation}\label{DNtm}
\Delta N_{\nu}^{\rm tot}=-0.8 \pm 0.4 \,\, ,
\end{equation}
implying a 3$\,\sigma$ range 
\begin{equation}
-2.0 < \Delta N_{\nu}^{\rm tot}<0.4 \,\, .
\end{equation}
Both results confirm our previous conclusion for which $\Delta N_{\nu}^{\rm tot}$
as high as $1$ is highly disfavoured, thus ruling out all four neutrino mixing models
{\em in the case of negligible neutrino asymmetries} \cite{main}.
However now both results seem to point out, at $2\,\sigma$, to a negative value of 
$\Delta N_{\nu}^{\rm tot}$, suggesting the presence of non standard BBN effects. 
We can also update the estimation of $\Delta N_{\nu}^{\rho}$ using
the new $\eta$ measurement from CMB and a new primordial Deuterium
abundance measurement \cite{Kirkman},
$(D/H)^{\rm exp}=(2.78^{+0.44}_{-0.38})\times 10^{-5}$, finding
$(\Delta N^{\rho}_{\nu})^{BBN}=0.7\pm 2.1$.
As already anticipated in \cite{main},
the error has been highly reduced by the great improvement in the
$\eta$ determination from CMB and it is now dominated by the error on $D/H$.
However, differently from the determination of $\Delta N_{\nu}^{\rm tot}$ 
from $Y_p^{\rm exp}$, better future determinations of $\eta$ (for example 
from new WMAP data or from Planck)
can still further reduce the error on $(\Delta N_{\nu}^{\rho})^{\rm BBN}$ 
from the current $2.1$ down to $1.5$.
  It is interesting to note that the value from BBN 
is comparable to the direct determination from CMB. In
\cite{pastor}, combining the WMAP data with the 2dF redshift survey and using
the value on the Hubble constant from the HST Key Project, $h=0.72\pm 0.08$ \cite{HST},
the authors find $(\Delta N^{\rho}_{\nu})^{CMB}=0.5^{+1.8}_{-0.9}$.
Assuming that, between the nucleosynthesis and the recombination time, the quantity
$\Delta N_{\nu}^{\rho}$ does not change
\footnote{See \cite{main} and \cite{Bowen} for discussions and examples in which 
$(\Delta N_{\nu}^{\rho})^{CMB}\neq (\Delta N_{\nu}^{\rho})^{BBN}$.}
and thus that 
$(\Delta N_{\nu}^{\rho})^{BBN}=(\Delta N_{\nu}^{\rho})^{CMB}$, one can then combine the two
values. We will still assume gaussian errors for a qualitative
estimation
\footnote{From the likelihood distribution given in \cite{pastor},  
this does not seem to be a very good approximation at values larger than the central one,
while it is reasonably good for smaller values.} and in this way
we find a CMB-Deuterium combined value
\begin{equation}\label{DNrho}
(\Delta N_{\nu}^{\rho})^{CMB+D/H}\simeq 0.6^{+1.4}_{-0.8}\,\, .
\end{equation}
In this way we get a much more stringent $2\,\sigma$ (3$\sigma$) range:
\begin{equation}
-(1.8)\,1.0 \lesssim (\Delta N^{\rho}_{\nu})^{CMB+D/H}\lesssim 3.4\,(4.8) \,\, .
\end{equation}

\section{Possible scenarios}

These new results show that deviations from Standard BBN,
if they exist, are small. This means that Standard BBN is in any case,
in first approximation, a very good description of all data. This result is mainly
due to the fact that the Deuterium abundance is in very good agreement
with the CMB prediction.
At the same time the measured primordial Helium abundance, $Y_p$, suggests 
the possible presence of  small deviations whose detection is now possible mainly to the
great precision of CMB in measuring the baryon asymmetry. However, for
an assessment of such a hint, it will be necessary to reduce
the large sistematic uncertainties on $Y_p$ and it will be also necessary to investigate
even more accurately on the robustness of the $\eta$ determination from CMB. 
In the following we will assume
that such a hint is suggestive of non standard BBN effects and we will discuss
some possible scenarios that could explain these deviations.
An important role in our discussion is given by the 
quantity $\Delta N_{\nu}^{f_{\nu}}=\Delta N_{\nu}^{\rm tot}-\Delta N_{\nu}^{\rho}$.
From (\ref{DNrho}) and (\ref{DNtm}) we can estimate:
\begin{equation}\label{DNf}
\Delta N_{\nu}^{f_{\nu}}\simeq -1.4^{+0.9}_{-1.4}
\end{equation}

\subsection{Low expansion rate}

A minimal possible way to interpret the data 
is to assume that there is no effect due to electron neutrino distribution
distortions and thus $\Delta N_{\nu}^{f_{\nu}}=0$ or equivalently
$\Delta N_{\nu}^{\rm tot}=\Delta N_{\nu}^{\rho}$. 
 In this case one can combine the result (\ref{DNtm})
from $Y_p$ and the result (\ref{DNrho}) from Deuterium plus CMB, 
getting $(\Delta N_{\nu}^{\rho})^{CMB+D/H+Y_p}=-0.6^{+0.40}_{-0.35}$.
 This result would  suggest a negative value of $\Delta N_{\nu}^{\rho}$, 
mainly due to the low value of $Y_p$,
implying a highly non standard modification of the expansion rate
during the BBN time, more precisely a lower expansion rate. Usually 
the presence of new particle species would lead to a higher expansion rate and
therefore such a possibility should rely on some drastic change of the radiation
dominated picture during the BBN period. However note that, from the 
Eq. (\ref{DNf}), the measurements mildly favour a 
value $\Delta N_{\nu}^{f_{\nu}}\neq 0$ and so this scenario is mildly
disfavoured from the data (at almost 90\% c.l.). 

\subsection{Degenerate BBN}

A well known modification of the Standard BBN is to introduce neutrino chemical potentials
in the thermal distributions \cite{DBBN}, corresponding to have pre-existing neutrino
asymmetries or generated at temperatures $T\gtrsim 10\,{\rm MeV}$ by 
some unspecified mechanism. 
An electron neutrino chemical potential ($\xi_e=\mu_e/T$) would yield 
$\Delta N_{\nu}^{\rm tot}\simeq -16\,\xi_e$.
The observed $Y_p$ (cf. (\ref{Ypm})) would then be explained by having
\begin{equation}\label{chp}
\xi_e=0.05\pm 0.025 \,\, .
\end{equation}
It has been shown in \cite{dopa}, extending the results of \cite{smirnov}, 
that the existing information on neutrino mixing
makes possible to conclude that before  the onset of BBN arbitrary initial 
neutrino chemical potentials would be almost equilibrated in a way that
$\xi_{\nu}\simeq \xi_{\tau}\simeq \xi_e$.
The presence of chemical potentials would thus correspond to
\begin{equation}
\Delta N_{\nu}^{\rho}\simeq 3\,
\left[{30\over 7}\left({\xi_{e}\over\pi} \right)^2+
{15\over 7}\left(\xi_{e}\over\pi \right)^4\right] \simeq 
3\times 10^{-3} \ll \Delta N_{\nu}^{\rm tot} 
\end{equation}
Therefore in this scenario the expansion rate would be practically standard and the deviations  would entirely arise from non standard electron neutrino distribution. 

\subsection{Active sterile neutrino oscillations}

Let us assume now that at temperature $T\gg 10\,{\rm MeV}$ all neutrino asymmetries
are negligible, for example of the order of the baryon asymmetry. 
It has been shown in many papers \cite{many} that a small mixing betwen
active neutrinos and new light sterile neutrinos can generate ordinary neutrino asymmetries
and thus negative values of $\Delta N_{\nu}^{f_{\nu}}$
together with $\Delta N_{\nu}^{\rho}\geq 0$. In a simplified
two neutrino mixing the value of $\Delta N_{\nu}^{f_{\nu}}$
is highly dependent on the value of the parameter $\Delta m^2_{i s}=m^2_{s}-m^2_i$. 
Usually the possibility to introduce active-sterile neutrino oscillations 
was motivated by the LSND anomaly \cite{LSND}. However
an explanation of the LSND anomaly in terms of active-sterile neutrino
oscillations, compatible with the solar and atmospheric neutrino
data would yield, as already mentioned, $\Delta N_{\nu}^{\rm tot}=\Delta N_{\nu}^{\rho}\sim 1$
\cite{main} (see also \cite{abaz}). At the same time 
the new WMAP bound on the neutrino masses, $m_{i}\leq 0.23\,{\rm eV}$ \cite{WMAP},
is now also incompatible with such an explanation of the LSND anomaly
\cite{Murayama}, except for one constrained exception \cite{Giunti}.
The possibility to generate a negative $\Delta N_{\nu}^{\rm tot}$
requires a negative value of $\Delta m^2=m^2_s-m^2_i$ and very small mixing angles
($\sin^2 2\theta \ll 10^{-4}$ \cite{many,main}). Values of $m_i\leq 0.23\,{\rm eV}$
imply thus $|\Delta m^2_{i s}|\leq 5\times 10^{-2}\,{\rm eV}^2$.
In \cite{Robert} it was shown how such maximum value, together with 
very small mixing angles, would produce
$\Delta N_{\nu}^{f_{\nu}}\geq  -0.3$.  For an inverted full hierarchical case
the corresponding $|\Delta m^2_{i s}|\sim 10^{-2}\,{\rm eV}^2$ and in this
case $\Delta N_{\nu}^{f_{\nu}}\sim -0.13$. These values have to be considered
as maximal because in the reality one should consider a full multi-flavour mixing and, though
full calculations are still missing, one can expect that part of 
electron neutrino asymmetry is actually shared with the other two flavours. 
This means that the small effect could 
reconcile the observed $\eta_B$ from CMB only with high values of $Y_p$
(cf. (\ref{high})).
In a two neutrino mixing small positive values of $\Delta N^{\rho}_{\nu}$ are also possible,
for larger mixing angles, but this would go at expenses of $|\Delta N_{\nu}^{f_{\nu}}|$,
making it even smaller \cite{main}. Having more than one sterile neutrino flavour
would make possible to have $\Delta N_{\nu}^{f_{\nu}}\simeq -0.3$ and positive
$\Delta N_{\nu}^{\rho}$ but in this case the total $\Delta N_{\nu}^{\rm tot}$ would be
larger than $-0.3$. This possibility is however interesting, since it would be 
a way to distinguish active-sterile neutrino oscillations  from a degenerate 
BBN scenario. Another way would be the detection of the effects
of a possible formation of neutrino domains \cite{domains},
like inhomeogeneities in the primordial Deuterium abundance \cite{domains}
that would give rise to gravitational waves \cite{Grasso}.

\subsection{Degenerate BBN and class A 3+1 models} \label{trepuno}

This is an intriguing variation of the pure degenerate BBN scenario. 
Suppose there are both large chemical
potentials and also a mixing of new sterile neutrino flavors with the ordinary ones.
If the chemical potentials are of the order
given by the Eq. (\ref{chp}), then, even for maximal mixing, 
the sterile neutrino production prior the onset of BBN would 
be suppressed \cite{rayrob} and consequently
the final value of $\Delta N_{\nu}^{\rm tot}$
would be the same as in the degenerate BBN scenario, while $\Delta N_{\nu}^{\rho}$
can be in principle slightly higher because of a initial sterile neutrino
highly diluted abundance. In this way it is very 
interesting that, as already noted in \cite{main}, the cosmological bound on 
four neutrino mixing models can be evaded. 
Moreover this same conclusion applies also to the
(current WMAP or {\em any future one}) bound  on the sum of neutrino masses applied to so called
`class A 3+1' four neutrino mixing models.  
They are such that the highest mass eigenstate is 
almost coinciding with a new sterile neutrino flavour and separated from 
the three lighter ones, almost coinciding with the ordinary ones 
(see \cite{main} for references and details), by the LSND gap. In this case the
sterile neutrino contribution to the fraction $\Omega_{\nu}/\Omega_m$
would be negligible and the bound on the sum of neutrino masses
would apply only to the three active mass eigenstates whose 
total mass is the same as in ordinary three neutrino models.  
Note also that among four neutrino mixing models these are the  only ones to be still 
marginally consistent with neutrino mixing experiments \cite{Tortola}. 

\subsection{Decaying sterile neutrino}\label{decay}

CP symmetrical decays of a sterile neutrino with a mass $m_{\rm st}\gg m_i$ into
electron neutrino plus, for example, some unknown scalar, 
with a lifetime $\tau$, could yield a positive  
$\Delta N_{\nu}^{\rho}$ and at the same time a negative $\Delta N_{\nu}^{\rm tot}$ 
analogously to the 
decaying ${\rm MeV}-\nu_{\tau}$ mechanism of Hansen and Villante \cite{Villante}, 
but with some important differences. A sterile neutrino abundance  
produced before the quark-hadron phase transition, at temperatures 
$T\gg 100\,{\rm MeV}$, is necessarily highly diluted compared 
to that of ordinary neutrinos. This can be re-generated at a lower temperature 
$\sim 15\,{\rm MeV}\,(m_{\rm st}/{\rm eV})^{1/3}$, implying $m_{\rm st}\lesssim 1\,{\rm KeV}$, 
by a possible neutrino mixing with ordinary states 
(see \cite{main} for details and references). 
If the decaying temperature $T(t=\tau)$ 
is approximately comprised in a window $(0.5-5)\,{\rm MeV}$, elastic scatterings
can partially or totally kinetically equilibrate the excess of
produced electron neutrino and anti-neutrinos but its total number
cannot be totally destroyed by the partially or completely frozen 
annihilations before the freezing of neutron to proton ratio. 
This symmetric excess of electron neutrinos and antineutrinos would yield a negative 
$\Delta N_{\nu}^{\rm tot}$ that can agree with the value (\ref{DNtm}).
Note that since the decaying neutrinos are sterile, 
their decays would anyway occur {\em out-of-equilibrium} even though 
in the ultrarelativistic regime. An interesting possibility is the case that 
the sterile neutrino is the LSND neutrino of class A 3+1 models, 
with $m_{\rm st}\sim \sqrt{\Delta m^2_{LSND}}$, like for the degenerate BBN
scenario in III D. Now however $\Delta N_{\nu}^{\rho}\lesssim 1$, because
the decays can only partly destroy what generated by the mixing. 
This scenario clearly suffers from fine tuning between the lifetime of sterile neutrinos 
and the time window between freezing of $\nu_e$ annihilations and of the neutron to proton 
ratio.  A way to circumvent this problem is to allow decays to be CP asymmetrical in a way
to realize a sort of Fukugita-Yanagida leptogenesis at low temperatures. In this case 
it is enough that the life time is shorter than the $n/p$ freezing time  
($\sim 10\,{\rm sec}$). This case would be very similar
to the scenario IIID  but with $\Delta N_{\nu}^{\rho}\lesssim 1$, 
the exact value depending on the  mixing and on the life time. Note that like in IIID
the cosmological bound on the sum of neutrino masses would be also evaded,
as pointed out in \cite{weiler}.

\section{Conclusions}

In future years a better understanding of sistematic uncertainties
in the measured $Y_p$ could strengthen or disprove the hint of non standard BBN effects.
At the same time improved data from
CMB experiments should both be able to measure $\Delta N_{\nu}^{\rho}$
with a precision of $\sim 0.1$ \cite{Turner,Bowen}
and make even more robust and precise the determination of $\eta_B$. 
If the primordial Helium anomaly will be confirmed,
implying negative $\Delta N_{\nu}^{\rm tot}<0$, then a key quantity in discriminating
among different explanations is the difference 
$\Delta N_{\nu}^{\rm tot}-\Delta N_{\nu}^{\rho}$.
If this will prove to be not zero and negative, then low expansion rate
scenarios will be ruled out, as already mildly suggested from current data, 
and a scenario with large chemical potentials
would be a possible explanation if at the same time $\Delta N_{\nu}^{\rho}\sim
{\cal O}(10^{-3}-10^{-2})$ (maybe detectable in a very optimistic case \cite{Turner}). 
In the case that $|\Delta N_{\nu}^{\rm tot}|<0.3$,
then active-sterile neutrino mixing can be a viable explanation too and if this is 
also accompanied by a positive value of $\Delta N_{\nu}^{\rho}$, then it will be actually favoured, since degenerate BBN would be ruled out.   
 We also pointed out that the degenerate BBN scenario could receive support
from neutrino mixing experiments. This because large neutrino asymmetries
would make `class A 3+1' four neutrino mixing models a viable cosmological solution  
without any limitation from the bound on the sum of
the neutrino masses. In this case one should receive a confirmation of 
LSND from MiniBoone \cite{mini} that would realize a nice consistency between current cosmological data (but we need a better understanding of primordial Helium measurements) and 
neutrino mixing experiments. We would be left with the tough 
theoretical problem to understand the origin
of large neutrino asymmetries. Current knowledge excludes the nice possibility 
of active-sterile neutrino oscillations themselves, but maybe a further 
investigation could change such a conclusion, in particular considering that
full multi flavor active-sterile neutrino mixing calculations
are still missing and that the role of phases in three neutrino mixing has never been 
studied \cite{main}. One possibility is that the sterile neutrino decays
generate the needed electron neutrino distortions.
This case could explain the current central values
of $\Delta N_{\nu}^{\rm tot}$  (cf. (\ref{DNtm}) ) and $\Delta N_{\nu}^{\rho}$ 
(cf. (\ref{DNrho})) but only future more accurate determination 
will allow to distinguish among the different scenarios first of all between the
standard scenario and possible non standard ones.\\

\noindent
{\bf Acknowledgements.}
P.D.B. is supported by the EU network ``Supersymmetry
and the Early Universe" (HPRN-CT-2000-00152). He wishes to thank F.~Borzumati,
E.~Masso, A.~Pomarol and M.~Quiros for their questions that stimulated 
the content of \ref{trepuno} and the anonymous referee whose comments stimulated 
\ref{decay}. While editing the paper, 
the second version of \cite{Murayama} appeared with independent similar conclusions  
about the possible role of large neutrino asymmetries.

\end{document}